\documentclass[12pt]{article}
\usepackage{graphics}
\usepackage{amssymb}
\usepackage{amsmath}

\newcommand{\rf}[1]{(\ref{#1})}
\newcommand{\beq}{\begin{equation}}
\newcommand{\eeq}{\end{equation}}
\newcommand{\bea}{\begin{eqnarray}}
\newcommand{\eea}{\end{eqnarray}}

\newcommand{\e}{\mbox{e}}
\renewcommand{\d}{\mbox{d}}

\renewcommand{\l}{\lambda}

\renewcommand{\a}{\alpha}

\newcommand{\ep}{\varepsilon}

\newcommand{\ra}{\rangle}
\newcommand{\la}{\langle}
\newcommand{\prt}{\partial}

\newcommand{\cD}{{\cal D}}

\newcommand{\bx}{{\bar{x}}}

\newcommand{\sla}{\sqrt{\l}}

\begin{document}

\begin{center}
\vspace{24pt}
{ \large \bf Putting a cap on causality violations in CDT}

\vspace{30pt}

{\sl J. Ambj\o rn}$\,^{a,b}$
{\sl R. Loll}$\,^{b}$,
{\sl W. Westra}$\,^{c}$
and {\sl S. Zohren}$\,^{d}$

\vspace{48pt}
{\footnotesize

$^a$~The Niels Bohr Institute, Copenhagen University\\
Blegdamsvej 17, DK-2100 Copenhagen \O , Denmark.\\
{ email: ambjorn@nbi.dk}\\

\vspace{10pt}

$^b$~Institute for Theoretical Physics, Utrecht University, \\
Leuvenlaan 4, NL-3584 CE Utrecht, The Netherlands.\\
email: loll@phys.uu.nl, ambjorn@phys.uu.nl\\

\vspace{10pt}

$^c$~Department of Physics, University of Iceland, \\
Dunhaga 3, 107 Reykjavik, Iceland\\
email: willem@raunvis.hi.is

\vspace{10pt}

$^d$~Blackett Laboratory, Imperial College,\\
London SW7 2AZ, United Kingdom.\\
email: stefan.zohren@imperial.ac.uk\\
}
\vspace{96pt}
\end{center}


\begin{center}
{\bf Abstract}
\end{center}

The formalism of causal dynamical triangulations (CDT) provides
us with a non-perturbatively defined model of quantum gravity,
where the sum over histories includes only causal space-time
histories. Path integrals of CDT and their continuum limits have been 
studied in two, three and four dimensions.
Here we investigate a generalization of the two-dimensional CDT model,
where the causality constraint is partially lifted by introducing 
weighted branching points,
and demonstrate that the system 
can be solved analytically in the genus-zero sector.

\vspace{12pt}
\noindent

\vspace{24pt}

\newpage

\subsection*{Introduction}\label{intro}

The idea of CDT, by which we mean the definition of 
quantum gravity theory via causal
dynamical triangulations, is two-fold: firstly, 
inspired by earlier ideas in the continuum theory
\cite{tei1,tei2}, we insist, starting from space-times with a
Lorentzian signature, that only causal histories contribute to
the quantum gravitational path integral.
Secondly, we assume the presence of a global time-foliation.

The formalism of dynamical triangulations (DT) provides a simple regularization
of the sum over {\it geometries} by providing a grid of
piecewise linear geometries constructed from elementary building blocks
(these are $d$-dimensional simplices of identical size and shape 
if we want to construct a $d$-di\-men\-sio\-nal
geometry, see \cite{book,leshouches} for reviews).
The ultraviolet cut-off is given by the edge length of the building blocks.
The causal variant CDT also uses DT as the regularization of the path integral.
A detailed description of which causal geometries are included in the grid
can be found in references \cite{al,ajl5}.

We emphasize that the use of triangulations is merely
a technical regularization of the assumed underlying continuum theory,
in the same way a lattice can be used for regularizing
a quantum field theory. By no means do we presuppose that 
space-time is literally 
made out of little simplices. Some support for the existence of an
underlying (non-perturbative) continuum quantum field theory
in higher dimensions has been provided in \cite{ajl5}-\cite{ajl3d} 
and seems to 
be in qualitative agreement with independent analyses carried out using
the renormalization group \cite{reu1,reu2,reu3,litim,reu4}.

While the CDT model is defined as a sum over causal space-time histories 
(each with an appropriate weight), one can ask whether this 
causality constraint can 
be lifted. One motivation for introducing it was that unrestricted 
summation over space-time histories\footnote{even when keeping the space-time 
topology fixed} leads
to a dominance of highly singular configurations in dimensions $d > 2$, 
which prevents the existence of a physically meaningful continuum 
limit of the regularized lattice theory \cite{bielefeld,simon,simon2}. 
On the other hand, if one takes the point of view that a maximal number of 
possible fluctuations should be included in the path integral (while still
leading to a meaningful result), one may wonder whether it is possible to 
reintroduce (a subclass of) configurations into the sum over geometries
which correspond to metric structures with causality violations.
The question we would like to pose is whether this can be done in a controlled
manner -- using the additional time-slicing structure present in CDT -- 
which avoids the problems encountered previously by DT, corresponding to 
an unrestricted inclusion of all such configurations. 

Because of the ready availability of analytic tools and the existence of analytical 
solutions, we will in a first step analyze the situation in two dimensions.
In this context, the issue has been addressed previously in a 2d toy model,
focussing on the effects of including a class of minimal wormholes in CDT, 
which can be said to violate causality only mildly 
and are much less abundant than general wormholes \cite{lw0,lw1,lw2}.
In the present work, we will look at the genus-zero sector of a generalized model of 
two-dimensional CDT, which in principle allows for the inclusion of arbitrary 
space-time topologies, as well as 
``outgrowths", that is, the sprouting of baby universes,
and associates with them a weight depending on the 
gravitational coupling constant.\footnote{The
first analysis of a CDT 
model with local ``decorations" was made in \cite{fgk}.} 
A key observation is that requiring the propagator of the 
model to reduce to that of standard
CDT when the bare coupling is taken to 
zero uniquely fixes the scaling of this
coupling, leading to a continuum limit 
where branching processes occur, but are scarce
compared to the situation in DT. 

The following sections deal with 
explaining this scaling argument, and with analytically
computing the genus-0 propagator (or loop-loop amplitude) 
and corresponding disc amplitude.
The computation of higher-genus amplitudes 
in this framework is currently under way, 
and may open new perspectives on the issue of 
the sum over topologies in theories of
quantum gravity, which are only apparent in a formulation 
that has at least some memory
of the {\it Lorentzian} structure of space-time built in.
However, we have as yet no definite statements to make about the properties of
general higher-genus amplitudes, the summability of the genus expansion or a 
generalization of the model to higher dimensions.

\subsection*{Two-dimensional CDT and Euclidean quantum gravity}\label{cdt}

Two-dimensional quantum gravity is a wonderful playground for 
``quantum geometry'', understood as the statistical sum over
geometries. The reason for this is that the action is trivial
as long as we ignore topology changes (and even then
it is almost trivial). One can therefore use entirely geometric reasoning
to derive relations between or properties of ``Green's functions''\footnote{
An early example is the proof \cite{ad} that the string tension of
bosonic string theory (regularized using DT \cite{adf,adfo}) does not scale,
thus providing a simple geometric understanding of the 
impossibility of defining
bosonic string theory in target space dimensions larger than or equal to 2. 
Other applications in non-critical string theory can be found in 
\cite{adj,durhuus}.}.
In this context it is convenient to study the proper-time 
``propagator'', namely,
the amplitude of geometries with two space-like boundaries
separated by a proper time (or geodesic distance) $t$. Although
the proper-time propagator is a special amplitude, it has the virtue that other
amplitudes, like the disc or cylinder amplitudes,
can be calculated from it \cite{kawai0,aknt,gk,aw,al}.
When the path integral representation of this propagator
is defined in the Lorentzian domain, using CDT, we can associate
with each of the causal, piecewise
linear Lorentzian space-time geometries a unique Euclidean geometry. 
After this rotation we perform the sum over the Euclidean geometries 
thus obtained.
The sum is now different from the usual Euclidean
sum over geometries, since it extends only over a strict 
subset of all Euclidean 
configurations, leading to an alternative quantization
of 2d quantum gravity (CDT). In the end, we can rotate back the propagator
from Euclidean to Lorentzian proper time 
if needed. In the remainder of this article we will stay in the 
Euclidean regime, as defined above. 

For ease of presentation, we will in the following use a continuum notation.
A derivation of the continuum expressions from the regularized
(lattice) expressions can be found in \cite{al}.
We will assume that space-time has the topology $S^1\times [0,1]$.
After rotation to Euclidean signature, the action is
\beq\label{2.a}
S[g_{\mu\nu}] = \l \int \d^2 \xi  \sqrt{\det g_{\mu\nu}(\xi)} +
x \oint \d l_1 + y\oint \d l_2,
\eeq
where $\l$ is  the cosmological constant, $x$ and $y$ are two so-called boundary
cosmological constants, $g_{\mu\nu}$ is the metric of a geometry
of the kind described above, and the line integrals refer to
the lengths of the in- and out-boundaries induced by $g_{\mu\nu}$.
The propagator $G_\l(x,y;t)$ is defined by
\beq\label{2.a0}
G_\l (x,y;t) =
\int \cD [g_{\mu\nu}] \; e^{-S[g_{\mu\nu}]},
\eeq
where the functional integration is over all {\it causal} geometries $[g_{\mu\nu}]$
such that the final boundary with boundary cosmological constant
$y$ is separated a geodesic distance $t$ from the initial boundary
with boundary cosmological constant $x$.\footnote{This class 
of geometries is difficult to define directly in a continuum, gauge-fixed formulation; what we
have in mind here is the continuum limit of the corresponding discrete sum.} 
Calculating the path integral \rf{2.a0} with the help of the CDT regularization and taking
the continuum limit as the side-length $a$ of the simplices goes
to zero leads to the equation \cite{al}
\beq\label{2.n1}
\frac{\prt}{\prt t} G_\l(x,y;t) = - 
\frac{\prt}{\prt x} \Big[(x^2-\l) G_\l(x,y;t)\Big],
\eeq
which is solved by\footnote{The
asymmetry between $x$ and $y$ is due to the convention that
the initial boundary contains a marked point. Symmetric expressions
where neither or both boundaries have
marked points can be found in \cite{alnr}.}
\beq\label{2.a3}
G_\l (x,y;t) = \frac{\bx^2(t,x)-\l}{x^2-\l} \; \frac{1}{\bx(t,x)+y},
\eeq
where $\bx(t,x)$ denotes the solution of the characteristic equation for \rf{2.n1}, namely,
\beq\label{2.3}
\frac{\d \bx}{\d t} = -(\bx^2-\l),~~~\bx(0,x)=x.
\eeq

Let $l_1$ denote the length of the initial and $l_2$ the
length of the final boundary. Rather than considering a situation
where the boundary cosmological constant $x$ is fixed, we will
take $l_1$ as fixed, and denote the corresponding propagator by
$G_\l (l_1,y;t)$, with similar definitions for $G_\l(x,l_2;t)$
and $G_\l(l_1,l_2;t)$. All of them are related by Laplace transformations,
for instance,
\beq\label{2.a5}
G_\l(x,y;t)= \int_0^\infty \d l_2 \int_0^\infty \d l_1\;
G_\l(l_1,l_2;t) \;\e^{-xl_1-yl_2},
\eeq
where the Laplace-transformed propagator obeys the composition rule
\beq\label{2.a6}
G_\l (x,y;t_1+t_2) = \int_0^\infty \d l \;
G_\l (x,l;t_1)\,G_\l(l,y;t_2). 
\eeq
Eq.\ \rf{2.a6} is the simplest example of the use of quantum 
geometry. While the property \rf{2.a6} is evident in the context of 
CDT where no baby universes are allowed, it is also 
true in Euclidean quantum gravity (where there is no such 
constraint), if one defines the distance between the initial and final 
loop appropriately \cite{kawai0,watabiki}.

\begin{figure}[t]
\centerline{\scalebox{0.6}{\rotatebox{0}{\includegraphics{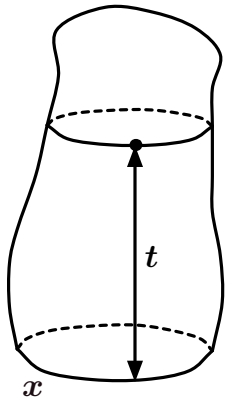}}}}
\caption[fig1]{{\small 
Graphical representation of relation \ref{2.50}: differentiating the 
disc amplitude $W_\l(x)$ (represented by the entire figure) with respect to
the cosmological constant $\lambda$ corresponds to marking a point somewhere inside
the disc. This point has a geodesic distance $t$ from the initial loop.
Associated with the point one can identify a connected curve of length $l$, 
all of whose points
also have a geodesic distance $t$ to the initial loop. This loop can now be
thought of as the curve along which the lower part of the figure
(corresponding to the loop-loop propagator $G_\lambda (x,l;t)$) is glued
to the cap, which itself is the disc amplitude $W_\l(l)$.}}
\label{fig1}
\end{figure}

Another, slightly more complicated example is 
illustrated graphically by Fig.\ \ref{fig1},
which implies the functional relation
\beq\label{2.50}
-\frac{\prt W_\l(x)}{\prt \l} =
 \int_0^\infty \d t \int_{0}^{\infty} 
\d l\ G_\l (x,l;t)\, l W_\l(l).
\eeq  
It encodes the following: let $W_\l(l)$ denote the disc amplitude,
i.e.\ the Hartle-Hawking amplitude with a fixed boundary length $l$, 
and $W_\l(x)$ the corresponding 
Laplace-transformed amplitude where $x$ is a fixed boundary 
cosmological constant. Differentiation with respect to the cosmological 
constant $\l$ means marking a point in the bulk, as shown in the 
figure.  Each configuration appearing in the 
path integral has a unique decomposition into a cylinder of proper-time extension $t$,
(where the proper time is defined as the geodesic distance of the marked point
to the boundary),
and the disc amplitude itself, as summarized in eq.\ \rf{2.50}.

Starting from a regularized theory with a cut-off $a$, it was shown in \cite{al} 
that there are two natural solutions to eq.\ \rf{2.50}.
In one of them, the regularized disc amplitude diverges with 
the cut-off $a$ and the geodesic distance $t$ scales canonically
with the lattice spacing $a$ according to
\bea\label{2.51}
W_{reg} &\xrightarrow[a\to 0]{}& a^{\eta}\, W_\l(x),~~~~\eta < 0, \\
t_{reg} &\xrightarrow[a\to 0]{}&  t/a^\ep,~~~~\ep =1.
\label{2.51a}
\eea
In the other, the scaling goes as
\bea\label{2.52}
W_{reg} &\xrightarrow[a\to 0]{}& {\rm const.} +a^{\eta}\, W_\l(x), 
~~~~\eta=3/2\\
t_{reg}& \xrightarrow[a\to 0]{}&  t/a^\ep,~~~~~\ep=1/2,
\label{2.52a}
\eea
where the subscript ``{\it reg}" denotes the regularized quantities in the discrete
lattice formulation.
The first scaling \rf{2.51}-\rf{2.51a}, with $\eta =- 1$, is encountered in CDT,
while the second scaling \rf{2.52}-\rf{2.52a} is realized in Euclidean
gravity, i.e.\ Liouville gravity or gravity defined from matrix models.

As demonstrated in \cite{al}, it is possible to treat both models simultaneously.
Allowing for the creation of baby universes during the ``evolution''
in proper time $t$ (by construction, a process forbidden in CDT) leads
to a generalization of \rf{2.n1}, namely,
\beq\label{2.53}
 a^{\ep}\frac{\prt}{\prt t} G_{\l,g}(x,y;t) = 
- \frac{\prt}{\prt x} \Big[\Big(a(x^2-\l)+2 g\, a^{\eta-1} W_{\l,g}(x)\Big) 
G_{\l,g}(x,y;t)\Big],
\eeq
where we have introduced a new coupling constant $g$, associated
with the creation of baby universes, and also made the additional dependence
explicit in the amplitudes.
In \cite{al} it was noted that for $g =1$, that is,
viewing this creation as a purely
geometric process\footnote{By this we mean that each distinct geometry (distinct in the
sense of Euclidean geometry) appears with equal
weight in the sum over two-dimensional geometries.}, one obtains 
Euclidean quantum gravity. This happens because according
to \rf{2.51} and \rf{2.52}, we have either $\eta= -1$, which is 
inconsistent with \rf{2.53}, or we have from \rf{2.52} that 
$\eta = 3/2$ and thus $\ep = 1/2$, which is consistent with \rf{2.52}.
On the other hand, setting $g = 0$, thereby forbidding the creation of 
baby universes, leads of course back to \rf{2.n1}.

\begin{figure}[t]
\centerline{\scalebox{0.5}{\rotatebox{0}{\includegraphics{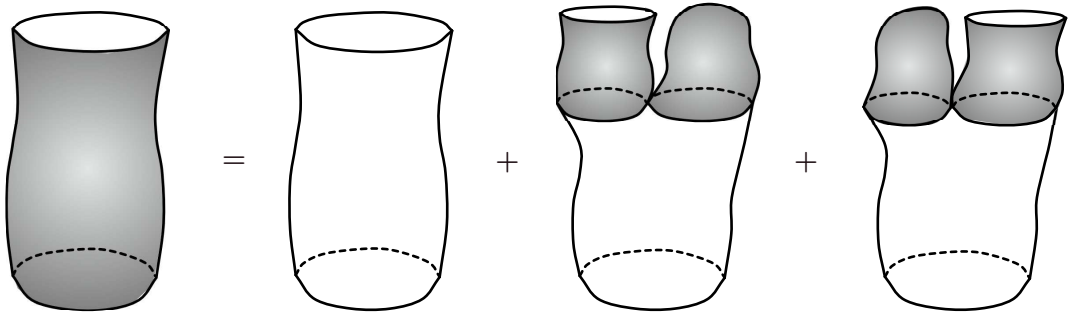}}}}
\caption[fig2]{{\small 
In all four graphs, the geodesic distance from the final to the initial 
loop is given by $t$. Differentiating
with respect to $t$ leads to eq.\ \rf{2.55}. Shaded parts of graphs represent
the full, $g_s$-dependent propagator and disc amplitude, and non-shaded 
parts the CDT propagator.}}
\label{fig2}
\end{figure}

In a non-trivial extension of previous work, we will now allow for the possibility that the 
coupling $g$ becomes a 
non-constant function $g = g(a)$ of the cut-off $a$. A geometric interpretation of this
assignment will be given in the discussion section below. 
Since we are 
interested in a theory  which smoothly recovers CDT in the limit as $g \to 0$, it 
is natural to assume that $\eta = - 1$,  like in CDT. Consequently, the 
only way to obtain a non-trivial consistent equation is 
to assume that $g$ scales to zero with the cut-off $a$ according to
\beq\label{2.54}
g = g_s a^3,
\eeq
where $g_s$ is a coupling constant of mass dimension three, which is 
kept constant when $a \to 0$. With this choice, eq.\ \rf{2.53} is turned into
\beq\label{2.55}
 \frac{\prt}{\prt t} G_{\l,g_s}(x,y;t) = 
- \frac{\prt}{\prt x} \Big[\Big((x^2-\l)+2 g_s\;W_{\l,g_s}(x)\Big) 
G_{\l,g_s}(x,y;t)\Big].
\eeq
The graphical representation of eq.\ \rf{2.55} (or \rf{2.53} for $g\not= 0$) 
is shown in Fig.\ \ref{fig2}.
Differentiating the integral equation corresponding to this 
figure with respect to the time $t$ one obtains \rf{2.55}.
The disc amplitude $W_{\l,g_s}(x)$ is at this stage unknown. 

Note that one could in principle have considered an a priori more general branching process, 
where more than one baby universe is allowed to sprout at any given time step $t$. 
However, one observes from the scaling relation \rf{2.54} that the corresponding extra terms in 
relation \rf{2.53} would be suppressed by higher orders of $a$ and therefore play no role in the continuum limit.

In the next section we will show that quantum geometry, in the sense defined 
above, together with the requirement of recovering standard CDT in the
limit as $g_s\rightarrow 0$,
uniquely determines the disc amplitude and thus 
$G_{\l,g_s}(x,y;t)$.

\subsection*{The disc amplitude}\label{disc}

\begin{figure}[t]
\centerline{\scalebox{0.55}{\rotatebox{0}{\includegraphics{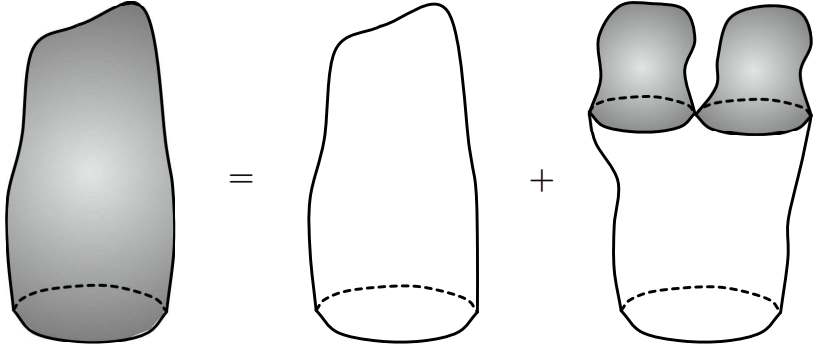}}}}
\caption[fig3]{{\small Graphical illustration of eq.\ \rf{3.2}. Shaded
parts represent the full disc amplitude, unshaded parts the CDT disc
amplitude and the CDT propagator. 
}}
\label{fig3}
\end{figure}

The disc amplitude of CDT was calculated in \cite{al,ackl}.
In \cite{al} it was determined directly by integrating $G_\lambda(l_1,l_2=0;t)$
over all times. This decomposition is unique, since by assumption 
$t$ is a global time and no baby universes can be created. In \cite{ackl}
it was shown that it could also be obtained from Euclidean quantum gravity
(matrix model results) by peeling off baby universes in a systematic way.
By either method one finds
\beq\label{3.1}
W_\l (x) = \dfrac{1}{x+\sla}
\eeq
for the disc amplitude as function of the boundary cosmological constant $x$.
In the present, generalized case we allow for baby universes, leading to a
graphical representation of the decomposition of the disc amplitude as 
shown in Fig.\ \ref{fig3}.
It translates into the equation 
\beq\label{3.2}
W_{\l,g_s} (x) = W_{\l,g_s} ^{(0)}(x) + g_s\int\limits_0^\infty \d t \int\limits_0^\infty \d l_1 \d l_2  \;
(l_1+l_2) G^{(0)}_{\l,g_s}  (x,l_1+l_2;t) W_{\l,g_s} (l_1)W_{\l,g_s} (l_2)
\eeq
for the full propagator $W_{\l,g_s}(x)$, where we have introduced 
a superscript $(0)$ to indicate the CDT amplitudes, that is,
\beq
W_{\l,g_s} ^{(0)}(x)\equiv W_{\l,g_s=0} (x)=W_\l(x),\label{relabel}
\eeq 
and similarly for $G^{(0)}_{\l,g_s}$, quantities which were defined in eqs. \rf{3.1} and \rf{2.a3}
respectively.
The integrations in \rf{3.2} can be performed, yielding
\beq\label{3.3}
W_{\l,g_s}(x) = \dfrac{1}{x+\sla} +\dfrac{g_s}{x^2-\l}\Big( W_{\l,g_s}^2(\sla)-W_{\l,g_s}^2(x)\Big).
\eeq
Solving for $W_{\l,g_s}(x)$ we find
\beq\label{3.4}
W_{\l,g_s}(x) = \frac{-(x^2-\l) + \hat{W}_{\l,g_s}(x)}{2g_s},
\eeq
where we have defined
\beq
\hat{W}_{\l,g_s}(x) = \sqrt{(x^2-\l)^2 +4g_s\Big(g_s W^2_{\l,g_s}(\sla) + x-\sla\Big)}.
\label{3.4a}
\eeq
The sign of the square root is fixed by the requirement that 
$W_{\l,g_s}(x) \to W_\l(x)$ for $g_s \to 0$, and $W_{\l,g_s}(x)$ is determined
up to the value $W_{\l,g_s}(\sla)$. We will now show that this value is also 
determined by consistency requirements of the quantum geometry. If we insert the 
solution \rf{3.4} into eq.\ \rf{2.55} we obtain
\beq\label{3.5}
 \frac{\prt}{\prt t} G_{\l,g_s}(x,y;t) = 
- \frac{\prt}{\prt x} \Big[\hat{W}_{\l,g_s}(x)\, G_{\l,g_s}(x,y;t)\Big].
\eeq
In analogy with \rf{2.a3} and \rf{2.3}, this is solved by
\beq\label{3.6}
G_{\l,g_s} (x,y;t) = \frac{\hat{W}_{\l,g_s}(\bx(t,x))}{\hat{W}_{\l,g_s}(x)} \; \frac{1}{\bx(t,x)+y},
\eeq
where $\bx(t,x)$ is the solution of the characteristic equation for \rf{3.5}, 
\beq\label{3.7}
\frac{\d \bx}{\d t} = -\hat{W}_{\l,g_s}(\bx),~~~\bx(0,x)=x,
\eeq
such that
\beq\label{3.8}
t = \int^x_{\bx(t)} \dfrac{\d y}{\hat{W}_{\l,g_s}(y)}.
\eeq
Physically, we require that $t$ can take values from 0 to $\infty$, as opposed to just in a
finite interval. From expression \rf{3.8} for $t$ this is 
only possible if the polynomial under the square root in the defining equation 
\rf{3.4} has a double zero, which fixes the function $\hat{W}_{\l,g_s}(x)$ to
\beq\label{3.9}
\hat{W}_{\l,g_s}(x) = (x-\a)\sqrt{(x+\a)^2-2g_s/\a},
\eeq
where 
\beq\label{3.9a}
\a = u\sla, ~~~u^3-u+\dfrac{g_s}{\l^{3/2}}=0.
\eeq
In order to have a physically acceptable $W_{\l,g_s}(x)$, 
one has to choose the solution to the third-order equation which is closest to 1. 
Quite remarkably, one can also derive \rf{3.9} from \rf{3.4} by demanding that 
the inverse Laplace transform $W_{\l,g_s}(l)$ fall off exponentially for large $l$.
In this region $W_{\l,g_s}(x)$ equals $W^{(0)}_{\l,g_s}(x)$ plus a convergent 
power series in the
dimensionless coupling constant $g_s/\l^{3/2}$.

One can check the consistency of the quantum geometry by noting that using 
\rf{3.6} in \rf{2.50} the integration can be performed to yield
\beq\label{3.10}
\dfrac{\prt W_{\l,g_s} (x)}{\prt \l} = \dfrac{W_{\l,g_s} (x) - W_{\l,g_s} (\a)}{\hat{W}_{\l,g_s}(x)},
\eeq
which is indeed satisfied by the solution \rf{3.4}.

\subsection*{The loop-loop amplitude}

We mentioned above that the loop-loop propagator can be regarded as a 
building block for other, more conventional ``observables" in 
2d quantum gravity. One of the most beautiful illustrations of this
and at the same time a non-trivial example of what we have called quantum 
geometry is the calculation in 2d Euclidean quantum gravity of the loop-loop 
amplitude from the loop-loop proper-time
propagator \cite{aknt}. The full loop-loop amplitude is obtained by summing 
over all Euclidean 2d geometries with two boundaries, without any particular 
restriction on the boundaries' mutual position. This amplitude 
was first calculated using matrix model techniques (for cylinder topology) \cite{ajm}. 

\begin{figure}[t]
\centerline{\scalebox{0.55}{\rotatebox{0}{\includegraphics{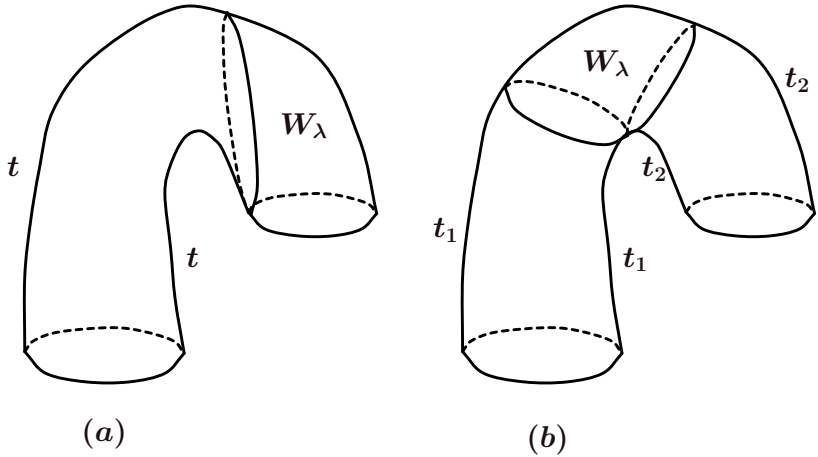}}}}
\caption[fig4]{{\small Two different ways of decomposing
the loop-loop amplitude into proper-time propagators and a disc amplitude.
Two points touch in the disc amplitude $W$, pinching the boundary to a figure-8, 
which combinatorially 
implies a substitution $W_{\l,g_s} (l) \to l W_{\l,g_s}(l)$ in the formulas. The time variables
are related by $t_1+t_2 = t$.
}}
\label{fig4}
\end{figure}

To appreciate the underlying construction, consider a given geometry of cylindrical
topology.
Its two boundaries will be separated by a geodesic distance $t$, in the sense of
minimal distance of any point on the final loop to the initial loop. It follows that we can 
consider the geometry as composed of a cylinder where the entire final loop (i.e.\ {\it each}
of its points) has a distance $t$ from the initial one and a ``cap" related to the
disc amplitude, as illustrated in Fig.\ \ref{fig4}(a). 
One can now obtain the
loop-loop amplitude by integrating over all $t$ and all gluings of the cap
(we refer to \cite{aknt} for details). An intriguing aspect of the 
construction is that the decomposition of a given geometry into cylinders
and caps is not unique. One can choose another decomposition
consisting of two cylinders of length $t_1$ and $t_2$, with $t_1+t_2 = t$,
joined by a cap, as illustrated in Fig.\ \ref{fig4}(b). 
As shown in \cite{aknt}, the end result is indeed independent of this decomposition.

The whole construction can be repeated for our new, generalized CDT model, in this 
way {\it defining} a loop-loop amplitude. More precisely, although an exact equality of
amplitudes corresponding to different
decompositions like those depicted in Fig.\ \ref{fig4}(a) and (b) is not 
immediately obvious at
the level of the triangulations of the discretized theory\footnote{because of the 
different arrangements of the proper-time slicings}, the continuum ansatz \rf{4.1} 
below is self-consistent, in the sense that it leads to a non-trivial symmetric 
expression for the amplitude with a well-defined $g_s\rightarrow 0$ limit. 
The algebra is similar to that of \cite{aknt}.

We will denote the loop-loop amplitude by $G_{\l,g_s}(x,y)$, and its
Laplace transform by $G_{\l,g_s}(l_1,l_2)$, related in the same way 
as was discussed for the loop-loop propagator (c.f. 
eq.\ \rf{2.a5} and the discussion leading up to it). 
The integral equation corresponding
to Fig.\ \ref{fig4} is given by 
\beq\label{4.1}
G_{\l,g_s}(l_1,l_2) = \int_0^\infty \d t \int_0^\infty \d l \; 
G_{\l,g_s}(l_1,l;t) l W_{\l,g_s}(l+l_2).
\eeq
Laplace-transforming eq.\ \rf{4.1}, the integrals can be performed
using  eqs.\ \rf{3.6}-\rf{3.9}. After some non-trivial algebra one obtains
\beq\label{4.2}
G_{\l,g_s}(x,y) = \dfrac{1}{f(x)f(y)}\dfrac{1}{4g_s}
\left( \dfrac{[(x+\a)+(y+\a)]^2}{(f(x)+f(y))^2}-1\right),
\eeq
where we are using the notation
\beq\label{4.3}
f(x) = \sqrt{(x+\a)^2-2g_s/\a} = \hat{W}_{\l,g_s}(x)/(x-\a).
\eeq
In the limit $g_s \to 0$ one finds
\beq\label{4.4}
G^{(0)}_{\l,g_s}(x,y) = \dfrac{1}{2\sla (x+\sla)^2(y+\sla)^2},
\eeq
a result which could of course also have been obtained directly from
\rf{4.1} using \rf{2.a3}, \rf{2.3} and \rf{3.1}.
We note that the corresponding expression in the case of Euclidean 2d quantum gravity is
given by 
\beq\label{4.5}
 G^{(e)}_\l(x,y) = \dfrac{1}{2h(x)h(y)(h(x)+h(y))^2},~~~h(x)=\sqrt{x+\sla},
\eeq
which can be obtained from expressions similar to \rf{3.6}-\rf{3.9},
only with $\hat{W}_{\l,g_s}(x)$ replaced by the Euclidean disc amplitude
\beq\label{4.6}
W_\l^{(e)}(x) = (x-\sla/2)\; h(x).
\eeq
We observe a structural similarity between \rf{4.2} and \rf{4.6}, with the function $f(x)$ having the 
same relation to $\hat{W}_{\l,g_s}(x)$ as $h(x)$ has to $W^{(e)}_\l(x)$.
The existence of well-defined, symmetric expressions for the unrestricted loop-loop
amplitudes in our generalized CDT model (at genus 0) and thus in standard two-dimensional
CDT, formulas \rf{4.2} and \rf{4.4}, gives strong support to the claims that (i) the proper-time
propagator does indeed encode the complete information on the quantum-gravitational system,
and (ii) following the arguments given in \cite{aknt} concerning the decomposition invariance of
the loop-loop amplitude (c.f. Fig.\ \ref{fig4}), the continuum theory is diffeomorphism-invariant.

\subsection*{Discussion}

The generalized CDT model of 2d quantum gravity we have defined in this
paper is a perturbative deformation of the original model in the sense
that it has a convergent power expansion of the form
\beq\label{5.1}
W_{\l,g_s}(x) = \sum_{n=0}^\infty c_n (x,\l) \left( \dfrac{g_s}{\l^{3/2}}\right)^n
\eeq 
in the dimensionless 
coupling constant $g_s/\l^{3/2}$. This implies in particular that 
the average number $\langle n\rangle$ of  ``causality violations" in a 
two-dimensional universe
described by this model is finite, a property already observed in previous 2d models with
topology change \cite{lw0,lw1,lw2}.
The expectation value of the number $n$ of 
branchings can be computed according to
\beq\label{5.2}
\la n\ra = \dfrac{g_s}{W_{\l,g_s}(x)}\dfrac{\d W_{\l,g_s}(x) }{\d g_s},
\eeq
which is finite as long as we are in the range of convergence of $W_{\l,g_s}(x)$.
As already mentioned, this coincides precisely with the range where the
function $W_{\l,g_s}(x)$ behaves in a physically acceptable way, namely, 
$W_{\l,g_s}(l)$
goes to zero like exponentially in terms of the length $l$ of the boundary loop. 
The same is true for the other functions considered, namely, $G_{\l,g_s}(l_1,l_2;t)$ 
and $G_{\l,g_s}(l_1,l_2)$.

The behaviour \rf{5.2} should be contrasted with that in 2d Euclidean quantum gravity,
and is reflected in the different scaling behaviours \rf{2.51a} and \rf{2.52a} for the time $t$. 
These scaling 
relations show that the effective continuum ``time unit" in Euclidean quantum gravity is
much longer than in CDT, giving rise to infinitely many causality violations for a 
typical space-time history which appears in the path integral  
when the cut-off $a$ is taken to zero. This phenomenon 
was discovered in the seminal paper \cite{kawai0}.

As we have already mentioned in the introduction, the calculations presented here should 
be seen as pertaining to the genus-0 sector of a generalized CDT model, which also 
includes a sum over space-time topologies. Although we have not given a precise
definition of the higher-genus amplitudes in this paper, one would expect them to be finite
order by order. If the handles are as scarce as are the baby universes in the genus-0 
amplitudes, it might even be that the sum over all genera is uniquely defined. Whether 
or not this is so will clearly also depend on the combinatorics of allowed handle configurations.

In the context of higher-genus amplitudes, it is natural to associate each handle with a
``string coupling constant", because one may think of it as a process where (one-dimensional) 
space splits and joins again, albeit as a function of an intrinsic proper time, rather than the
time of any embedding space. An explicit calculation reveals that in the generalized CDT model
this process is related with a coupling constant $g_s^2$ \cite{wz}, which one may think of
as two separate factors of $g_s$, associated with the splitting and joining respectively.

How does the disc amplitude fit into this picture?
From a purely Euclidean point of view all graphs appearing in
Fig.\ \ref{fig3} have the fixed topology of a disc.
However, from a Lorentzian point of view, which comes with a notion of time, it is
clear that the branching of a baby universe is associated with a change of
the {\it spatial} topology, a
singular process in a Lorentzian space-time \cite{louko}.
One way of keeping track of this in a Wick-rotated, Euclidean picture is
as follows. Since 
each time a baby universe branches off it also has to end somewhere,  
we may think of marking the resulting ``tip" with a puncture. From a
gravitational viewpoint, each new puncture corresponds to a topology
change and receives a weight $1/G_N$, where $G_N$ is Newton's constant,
because it will lead to a change by precisely this amount in the two-dimensional 
(Euclidean) Einstein-Hilbert action
\beq\label{5.3}
S_{EH} = -\dfrac{1}{2\pi G_N} \int \d^2\xi \sqrt{g} R.
\eeq
Identifying the dimensionless coupling constant 
in eq.\ \rf{2.53} with $g(a)= e^{-1/G_N(a)}$, one can introduce a {\it renormalized}
gravitational coupling constant by
\beq\label{5.4}
\dfrac{1}{G_N^{ren}} = \dfrac{1}{G_N(a)}+\dfrac{3}{2}\ln \l a^2.
\eeq
This implies that the {\it bare} gravitational coupling constant $G_N(a)$ goes to
zero like $1/|\ln a^3|$
when the cut-off vanishes, $a \to 0$, in such a way that the product $\e^{1/G_N^{ren}}/\l^{3/2}$ is 
independent of the cut-off $a$. We can now identify 
\beq\label{5.5}
\e^{-1/G_N^{ren}} = g_s/\l^{3/2}
\eeq
as the genuine coupling parameter in which we expand.

This renormalization of the gravitational (or string) coupling constant is reminiscent 
of the famous double-scaling limit in non-critical string
theory\footnote{It is called the double-scaling limit since  
from the point of view of the discretized theory it involves a
simultanous renormalization of the cosmological constant $\l$ and 
the gravitational coupling constant $G_N$. In this article we have already
performed the renormalization of the cosmological constant. For details
on this in the context of CDT we refer to \cite{al}.}. 
In that case one also has $g_s \propto e^{-1/G_N^{ren}}$, the only difference
being that relation \rf{5.4} is changed to
\beq\label{5.6}
\dfrac{1}{G_N^{ren}} = \dfrac{1}{G_N(a)}+\dfrac{5}{4}\ln \l a^2,
\eeq
whence the partition function of non-critical string theory appears  
precisely as a function  of the dimensionless coupling 
constant $g_s /\l^{5/4}$.

\subsection*{Acknowledgments}

All authors acknowledge support by
ENRAGE (European Network on
Random Geometry), a Marie Curie Research Training Network in the
European Community's Sixth Framework Programme, network contract
MRTN-CT-2004-005616. R.L. acknowledges 
support by the Netherlands Organisation for Scientific Research (NWO) under their VICI 
program.

\end{document}